\documentclass[aps,prl,epsfig,floats,twocolumn,superscriptaddress,amssymb,amsmath,showpacs]{revtex4}

\usepackage{color}

\usepackage{graphicx}

\begin{document}

\title{Effect of Bending Anisotropy on the 3D Conformation of Short DNA Loops}

\author{Davood Norouzi}

\affiliation{Institute for Advanced Studies in Basic Sciences
(IASBS), Zanjan 45195, P.O. Box 45195-1159, Iran}

\author{Farshid Mohammad-Rafiee}

\affiliation{Institute for Advanced Studies in Basic Sciences
(IASBS), Zanjan 45195, P.O. Box 45195-1159, Iran}

\author{Ramin Golestanian}

\affiliation{Department of Physics and Astronomy, University of
Sheffield, Sheffield S3 7RH, United Kingdom}

\date{\today}

\begin{abstract}
The equilibrium three dimensional shape of relatively short loops of DNA is studied
using an elastic model that takes into account anisotropy in bending
rigidities. Using a reasonable estimate for the anisotropy, it is
found that cyclized DNA with lengths that are not integer multiples
of the pitch take on nontrivial shapes that involve bending out of
planes and formation of kinks. The effect of sequence inhomogeneity
on the shape of DNA is addressed, and shown to enhance the
geometrical features. These findings could shed some light on the
role of DNA conformation in protein--DNA interactions.
\end{abstract}


\maketitle

Interactions between DNA and proteins that cause deformations in the
structure of DNA are essentially ubiquitous during many life
processes inside cells \cite{Cell}. For example, in eukaryotes the
packing of DNA into nucleosome has been shown to lead to formation
of sharp bends \cite{Luger97}. DNA packing in a viral capsid
involves a high degree of confining and bending of viral genome
inside a volume with dimensions that are comparable to the DNA
persistence length \cite{Baker}. DNA is also deformed by proteins
during gene expression, when relatively short loops of DNA are
formed \cite{Halford-andothers}. It is known that the shape of DNA
matters to its interaction with proteins such as RNA polymerase
\cite{Hu-2006}, and that proteins locate their specific targets on
DNA \cite{Cell}. Therefore, it will be important to understand the
role of mechanical effects such as tension, torsion, or bending and
their couplings in determining the shape of DNA and their
corresponding potential roles in positioning strategies
\cite{Calladine}. In addition, most cases of short genomes and
plasmids have circular shapes in physiological conditions, which
suggests that the exact shape of a circular short segment of DNA
could be of significant biological implications \cite{Cell}.

Conformational properties of relatively short DNA segments have been
the subject of recent studies. These include experiments on loop
formation and measurement of the persistence length in different
scales \cite{Shore,Cloutier04,Vologodskii05,Yuan08}, theoretical
works on the probability of loop formation and efforts to interpret
the findings of the experiments
\cite{Shimada,ZC-j,Marko04,Ranjith05,Nelson05,Cocco05,Wiggins06,Spakowitz06,Popov07,Frey2007},
and molecular dynamics simulations \cite{lankas06,sarah}. It is
generally agreed that while modeling DNA as an isotropic elastic rod
works perfectly for length scales larger the DNA persistence length
($\sim 50$ nm), more elaborate models are needed to explain the
conformational properties of shorter segments of DNA. These models
should take account of various nonlinearities and structural
properties of DNA elasticity that appear when shorter segments are
subject to extreme constraints on bending and twisting
\cite{Calladine,sarah2}.

One of the nonlinear features that could affect the elasticity of
DNA is the anisotropy in the effective bending rigidities
corresponding to bending into the major and minor grooves. Such
anisotropic bending elasticity models have been considered in
studies of DNA segments of about 10 base pairs (bp) \cite{maha}, and
shown to predict formation of kinks and modulations in the curvature in 2D
\cite{Farshid-03-05}. The anisotropy has also been shown to be
responsible for some of the geometrical features observed in
nucleosomal DNA \cite{Farshid-PRL05}. However, it is generally
presumed to be unimportant when the DNA segment is long enough to
have a few full helical turns, and in particular for segments of
about 100 bp that have been the subject of recent controversy
\cite{ZC-j,Vologodskii05}.

\begin{figure}[b]
\includegraphics[width=0.8\columnwidth]{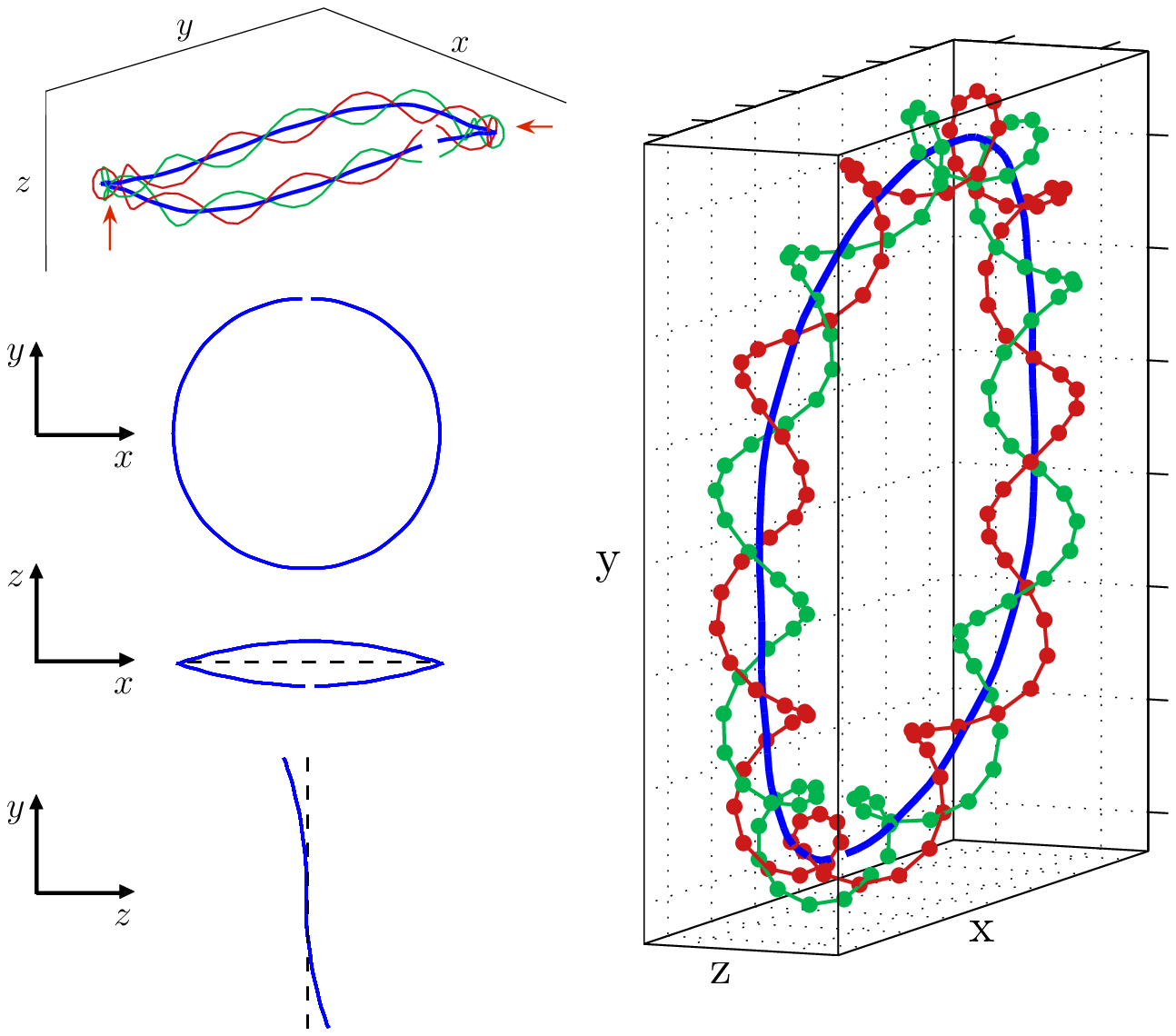}
\caption{(Color online) The ground-state conformation for a 94 bp
DNA loop with bending anisotropy ($A_1=75$ nm and $A_2=37$ nm). The
centerline of DNA is shown by a solid (blue) line. In the right
panel, the position of each base is shown by solid circles and the
two strands of DNA are shown in different colors. The positions of
kinks are shown by two arrows in the top left panel. }
\label{fig1:3D}
\end{figure}

Here we study the three dimensional shape of short fragments of
looped DNA using an elastic model that includes anisotropic bending
rigidities. We show that unlike the commonly accepted picture, the
anisotropy has a significant effect on the shape of DNA segments of
about 100 bp. We find that the equilibrium shape of DNA loops could
involve large changes in the local writhe and/or in the local
curvature, which we generally call kinks. Figure \ref{fig1:3D} shows
an example of a conformation with kinks (in the writhe). We also
examine the effect of sequence dependence on the actual shape by
incorporating varying bending rigidities within the region of
permissible values found in recent molecular dynamics simulations.
The sequence causes the DNA to soften/harden at some places, which
offers the structure the possibility of lowering the overall energy
by undergoing nontrivial deformations. The effect of the
sequence-dependent conformations are probed using the distributions
of local curvature, twist, and writhe, in the equilibrium shape.

To study the structure of looped DNA, we consider a simple model in
which the molecule is represented as an anisotropic elastic rod. The
rod is parametrized by the arclength $s$, and at each point an
orthonormal basis is defined with the unit vectors $\hat{e}_1(s)$,
$\hat{e}_2(s)$, and $\hat{e}_3(s)$, where $\hat{e}_1$ corresponds to
the direction from the minor groove to the major groove.
The deformation of the double helix is characterized by the angular
strains $\Omega_{1,2}(s)$ corresponding to bending in the plain
perpendicular to $\hat{e}_{1,2}(s)$ and $\Omega_3(s)$ corresponding
to twist. The elastic energy for the deformation of DNA in units of
thermal energy ($k_{\rm B}T \equiv 1/\beta$) is written as
\cite{Marko94}
\begin{math}
\beta E=\frac{1}{2} \int_0^L d s \left[ A_1 \Omega_1^2 + A_2
\Omega_2^2 + C(\Omega_3 - \omega_0)^2 \right],
\end{math}
where $A_1$ and $A_2$ are the bending rigidities for the ``hard''
and ``easy'' axes of DNA cross section, $C$ is the twist rigidity,
and $\omega_0 = 1.85 \; {\rm nm}^{-1}$ is the intrinsic twist of
B-DNA.
In terms of Euler angles,
we have $\Omega_1=\dot{\phi} \sin \theta \sin \psi + \dot{\theta}
\cos \psi$, $\Omega_2=\dot{\phi} \sin \theta \cos \psi -
\dot{\theta} \sin \psi$, and $\Omega_3=\dot{\phi} \cos \theta +
\dot{\psi}$, where the dot denotes differentiation with respect to
$s$. In order to find the ground state of the system we should
minimize this functional subject to the constraints in the system.
For closed loops one of the constraints is to conserve the linking
number, which can be expressed in terms of the Euler angles as $2
\pi Lk = \int_0^L ds (\dot \phi + \dot \psi) = (\phi + \psi) |_0^L$
\cite{Rudnick99}. Another constraint in the problem is a global
vector constraint that guarantees a closed loop, and is written as
$\int_0^L ds \, \hat{e}_3(s) = 0$. We also note that our Lagrangian
is invariant under a translation by one half of the length of the
loop. Therefore, we only focus on half of the loop, and demand that
the tangent vectors at the two ends of the two halves are opposite
to each other to ensure continuity. There is another important issue
with regards to the linking number. In order to have a closed loop
the two end base pairs should meet each other in phase, which means
that the linking number for a closed loop is an integer. As a
result, we might have to underwind or overwind the molecule to
produce a loop. In the case of B-DNA, the pitch is nearly 10 bp
long, and for example, a 94 bp fragment of DNA can form a loop if it
is underwound by the spontaneous twist of 4 bp, or overwound by that
of 6 bp, before the ends are joined up. It is not clear {\em a
priori} which one is more favorable, and one should compare the
energies of both solutions to decide that. Therefore we solve the
Euler-Lagrange differential equations subject to the boundary
conditions of $\phi(0)=\psi(0)=0$, $\theta(0)=0$, $\theta(L/2)=\pi$,
and $\psi(L/2)=\pi n$, where $n$ is the linking number of the DNA or
the number of DNA turns. Furthermore, we set $\dot{\phi}(0)=0$ just
to fix the starting plane of the loop.

The equilibrium structure of the loop depends on the values of the
elastic constants.
While a direct
experimental determination of the anisotropic bending rigidities is
still lacking, a recent simulation suggests a range of values of
$A_1=47-76$ nm and $A_2=24-51$ nm, depending on the sequence of the
nucleotides \cite{lankas}. In order to examine the effect of the
anisotropy, here we consider a representative case of $A_1=75$ nm
and $A_2=37$ nm, and compare it with the isotropic case with
$A_1=A_2=50$ nm. Both of these choices will lead to a persistence
length of $A=50$ nm, which can be calculated via
$A^{-1}=\frac{1}{2}(A_1^{-1}+A_2^{-1})$ \cite{LanLif}. There are
several suggested values for the DNA twist rigidity in the range of
50-110 nm \cite{Lankas2000}, and we consider a representative value
of $C = 75$ nm.

\begin{figure}
\includegraphics[width=0.90\columnwidth]{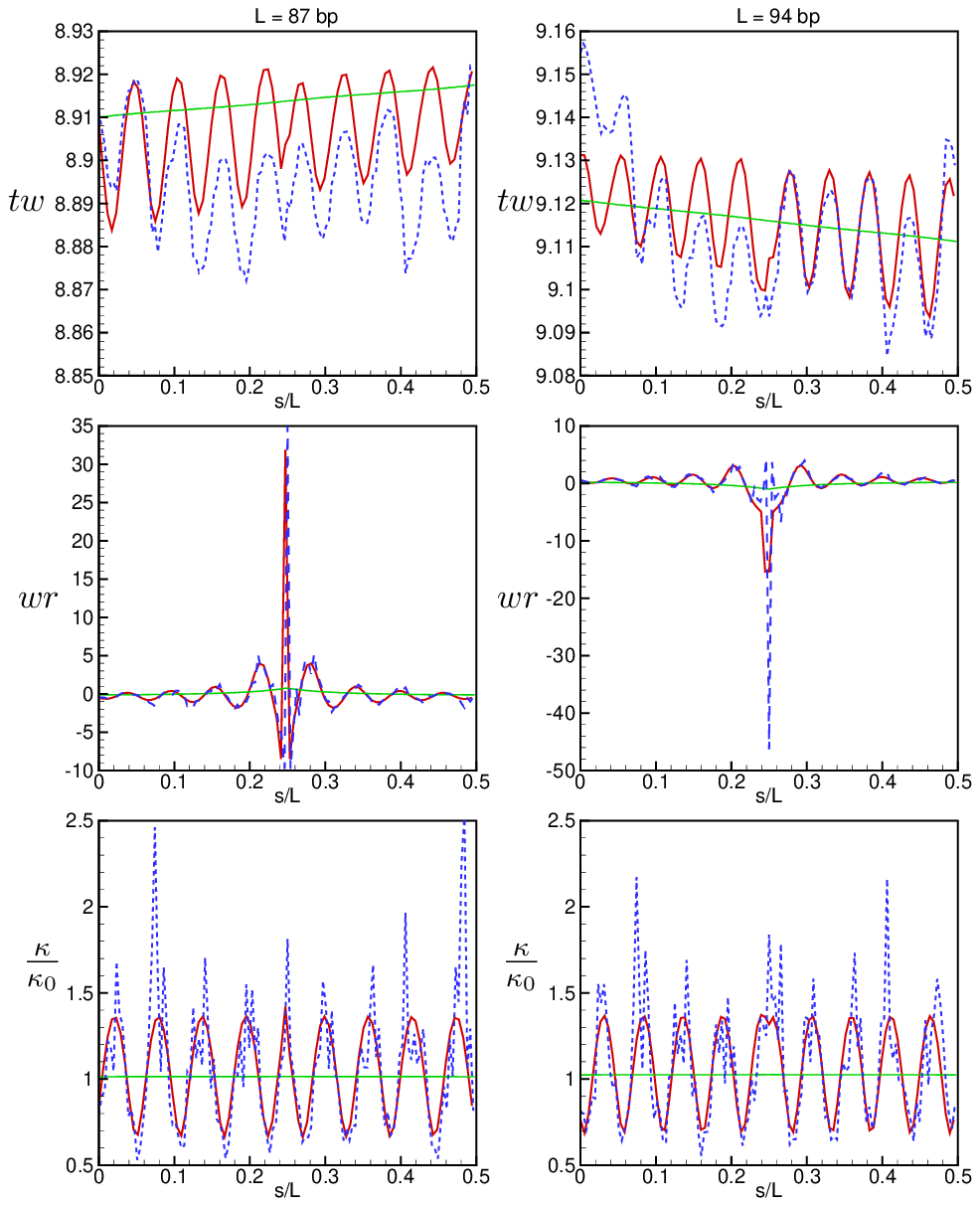}
\caption{(Color online) The local writhe $wr$, twist $tw$, and
curvature $\kappa$ for half of the looped DNA for the two lengths of
87 bp and 94 bp. The solid red lines correspond to $A_1=75$ nm and
$A_2=37$ nm, the dashed blue lines correspond to a Gaussian random
sequence centered around $A_1 = 75$ nm and $A_2 = 37$ nm with a
width of 7 nm in the effective persistence length (the mean value of
persistence length is 50 nm), and the light green lines correspond
to $A_1 = A_2 = 50$ nm.} \label{fig:local-curvature-writhe}
\end{figure}

In order to decrease the elastic deformation energy, the bent
anisotropic rod can explore non-planar configurations. To probe the
3D structure of the DNA, we calculate the values of the local
curvature, defined as $\kappa(s)=\sqrt{\Omega_1^2 +
\Omega_2^2}=\sqrt{\dot{\phi}^2 \sin^2 \theta + \dot{\theta}^2}$, the
local writhe $wr(s)=\frac{L}{2 \pi} \dot{\phi}(s) \left[1-\cos
\theta(s) \right]$, and the local twist $tw(s)=\frac{L}{2 \pi}
\left[ \dot{\phi}(s) \cos \theta(s) + \dot{\psi}(s) \right]$
\cite{Rudnick99}. In Fig. \ref{fig:local-curvature-writhe}, these
geometric measures are plotted for half of the DNA loop for two
lengths of 87 bp and 94 bp, for the anisotropic model. The curvature
is normalized with the uniform curvature of an isotropic untwisted circled rod
$\kappa_0 =2\pi/L$. Due to the anisotropy the local twist and the
local curvature modulate around their mean values. Note that an
increase in the curvature always coincides with a decrease in the
twist, and vice versa. The local writhe achieves nonzero values,
which indicates that the loop goes out of plane. The writhe could
actually become very large at a singular point (owing to large
values of $\dot{\phi}$), which corresponds to the presence of a kink
in the structure of the looped DNA, as shown in Fig. \ref{fig1:3D}.

Sequence inhomogeneity of DNA causes variations in the local bending
rigidities $A_1$ and $A_2$ along the DNA. The molecular dynamics
simulation by Lankas {\em et al.} \cite{lankas} suggests that
sequence dependence can cause a change of 10-20 $\%$ in the bending
rigidities locally. To see the effect of sequence on the shape of
short DNA loops, we consider a random sequence with a Gaussian
distribution around the anisotropic bending rigidities used above
with a width of 7 nm in the resulting persistence length around its
mean value of 50 nm. In Fig. \ref{fig:local-curvature-writhe}, the
effect of the Gaussian random sequence on the local shape parameters
has been shown. One can see that the curvature modulation in some
places have been sharpened, presumably because there are softer
places along the DNA available for accommodating the modulations
that are needed for overall equilibrium. While some of the bends are
sharp enough to be called kinks, the signature feature in the local
writhe seems to be also present in a system with a random sequence.
The bending kinks in the soft areas are accompanied by untwisting of
the helix. This behavior can be important for some biological
problems such as specific protein-DNA interactions, as it suggests
that depending on the sequence, there might be a possibility for
proteins to recognize the target location for an interaction. For
example, the so-called AT-boxes are known to be relatively easier
places for unwinding, and they are known to be the starting point
for some protein functions. Our results suggest that the anisotropy
can help the strands to be opened up more easily at softer sequences
and can facilitate the searching process of proteins along the DNA.

\begin{table}[b]
\caption{\label{tab:TwWr} Comparison of the geometric features and
energies for the isotropic (I), anisotropic (A), and sequence
dependent anisotropic (sd-A) models of DNA corresponding to the
examples of Fig. \ref{fig:local-curvature-writhe}. The mean
curvature along the DNA length is denoted by $\langle \kappa
\rangle$.}
\begin{ruledtabular}
\begin{tabular}{cccccccc}
model & $L {\rm (bp)}$ & $Tw$ & $Wr$
& $tw_{\rm min}$ & $\langle \kappa \rangle/\kappa_0$ & $\kappa_{\rm max}/ \kappa_0$ & $\beta E$ \\
\hline I & 87 & 8.91 & 0.09 & 8.90 & 1.01 & 1.03 & 36.6\\
A & 87 & 8.90 & 0.10 & 8.88 & 1.04 & 1.36 & 36.4\\
sd-A & 87 & 8.90 & 0.10 & 8.87 & 1.06 & 2.54 & 34.1\\
I & 94 & 9.12 & -0.12 & 9.11 & 1.02 & 1.05 & 36.2\\
A & 94 & 9.12 & -0.12 & 9.09 & 1.05 & 1.37 & 36.1\\
sd-A & 94 & 9.12 & -0.12 & 9.08 & 1.07 & 2.17 & 34.6\\
\end{tabular}
\end{ruledtabular}
\end{table}

It is also interesting to examine the effect of the anisotropy on
the total writhe $Wr$ and the total twist $Tw$ of DNA. In our
notation, they correspond to the average of their corresponding
local values, namely, $Wr=\frac{1}{L}\int_0^L wr(s) ds$ and
$Tw=\frac{1}{L}\int_0^L tw(s) ds $, where the local quantities
$wr(s)$ and $tw(s)$ are defined above. Table \ref{tab:TwWr} shows
the values for the total twist and writhe for two different DNA
lengths of 87 and 94 bp. One can check that the sum of the writhe
and the twist is constant for a given length, as required by White's
theorem \cite{White69}. Because in our examples the lengths of DNA
are not integer multiples of the pitch, the molecule should be
underwound or overwound to produce a loop. For the length of 94 bp,
the energy of the underwound DNA is smaller than the overwound one.
We note that the energy cost of changing the twist of the molecule
is reduced by the formation of writhe, and that the absolute value
of the total writhe becomes larger when the anisotropy increases. In
other words, the system tries to minimize the energy by combining
slight untwisting of the double strands with increased bending.
Implementing this strategy locally causes harmonious modulations in
the geometric measures of the DNA shape, as can be seen in Fig.
\ref{fig:local-curvature-writhe}. Table \ref{tab:TwWr} also shows
the extremal values of twist, and curvature for the homogeneous and
random sequence cases, which shows that randomness can allow for
significant enhancement in local geometrical features.

\begin{figure}
\includegraphics[width=0.65\columnwidth]{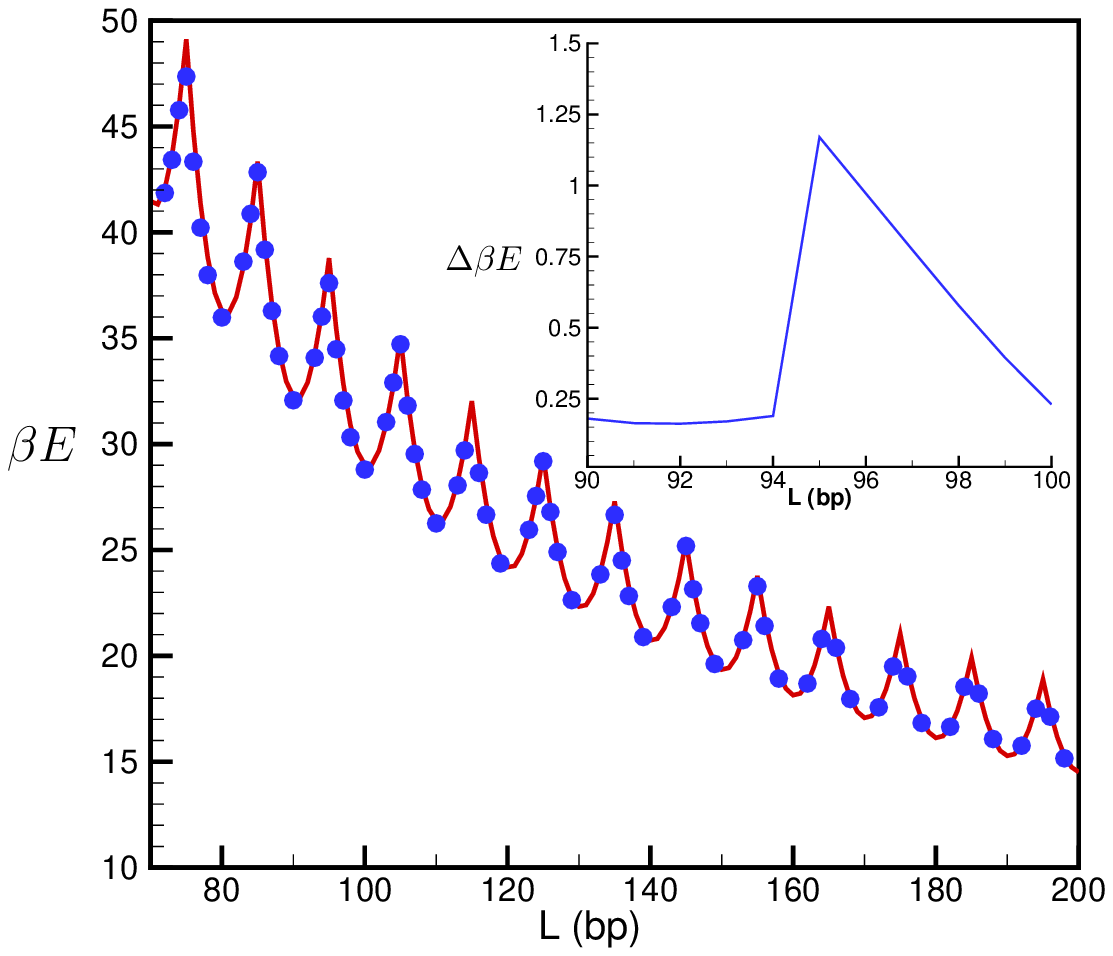}
\caption{(Color online) Energy of looped DNA as a function of its
length for the isotropic ($A_1=A_2=50$ nm, solid curve) and the
anisotropic ($A_1=75$ nm and $A_2=37$ nm, filled circles) models.
The inset shows the energy difference between two isotropic and
anisotropic looped DNA as a function of its length for the lengths
of 90-100 bp. } \label{fig:energy}
\end{figure}

Figure \ref{fig:energy} shows the total elastic energy of DNA loops
of different lengths for the two choices of the bending elastic
constants. The oscillations in the energy of the loop are because of
the additional twist deformation that is needed to make a full loop
(see above). We have calculated the energy for both over-twisted and
under-twisted configurations in every case, and used the lower value
in Fig. \ref{fig:energy}. For comparison, for the length of 94 bp
the energies of over- and under-twisted states are found as $41.8 \,
k_{\rm B} T$ and $36.1 \, k_{\rm B} T$, respectively. One can see
that the anisotropic system has a lower energy for all lengths, but
the difference (as compared to the thermal energy $k_{\rm B} T$) is
negligible. Introducing inhomogeneity in the bending rigidities (as
in the example shown in Fig. \ref{fig:local-curvature-writhe}) can
lower the energy of the system significantly, as Table
\ref{tab:TwWr} shows. As the energy difference between the isotropic
and anisotropic looped DNA is not large, we do not expect that
anisotropy without sequence dependency can change the $J$-factor
significantly. In the example given in Fig.
\ref{fig:local-curvature-writhe}, sequence disorder can reduce the
energy by $\sim 2 \, k_{\rm B} T$, which could increase the
$J$-factor value by a factor of $\sim 10$.

We find that both torsional and bending kinks will form in DNA loops
to reduce the elastic energy cost by taking advantage of bending and
torsion along the easy axis as much as possible. Atomistic modeling
studies using MD simulations support the presence of kinks along
highly deformed DNA \cite{lankas06}.
The kink corresponds to a concentrated region of high curvature
and/or torsion, and one wonders whether it might cause the hydrogen
bonds to break in that area. We find that the maximum bending energy
stored in a kink is $\sim 0.6 \; k_{\rm B} T$ per base pair for the
sequence dependent anisotropic model, which is much smaller than the
energy of the hydrogen bonds that is $\sim 10 \; k_{\rm B} T$ per
base pair. Finally, we note that the elastic energy of the
anisotropic model in the equilibrium conformation of the isotropic
model is $\sim 4 \, k_{\rm B}T$ higher than its own ground state
energy for the lengths used here. This means that the specific
geometrical features of the anisotropic looped DNA are robust, and
will not be blurred by thermal fluctuations.

In conclusion, we have studied the effect of bending anisotropy for
$\sim 100$ bp segments of looped DNA, and found that while the
anisotropy does not affect the overall elastic energy significantly,
it can have drastic effects on the shape in the form of the
emergence of kinks and unwindings. We also found a coupling between
DNA sequence, conformation, and its energy, which could be important
in the DNA cyclization probability and biological functions.

We are thankful to J. Gregory, M. Maleki, M. Kardar, G. I. Menon,
and M. Rao for valuable discussions.


\begin{thebibliography}{}


\bibitem{Cell}
B. Alberts \textit{et al.}, \textit{Molecular Biology of the Cell,
 4th ed.} (Garland, New York, 2002).

\bibitem{Luger97}
K. Luger \textit{et al.}, Nature (London) {\bf 389}, 251
(1997); T. J. Richmond and C. A. Davey, Nature (London) {\bf 423}, 145 (2003).

\bibitem{Baker}
T. S. Baker \textit{et al.}, Microbiol. Mol. Biol. Rev. {\bf 63},
862 (1999).

\bibitem{Halford-andothers}
S. E. Halford \textit{et al.}, Annu. Rev. Biophys.
Biomol. Struct. {\bf 33}, 1 (2004); R. Schleif, Annu. Rev. Biochem.
{\bf 61}, 199 (1992); S. Adhya, Annu. Rev. Genet. {\bf 23}, 227
(1989); H. G. Garcia {\it et al.}, Biopolymers {\bf 85}, 115 (2006).

\bibitem{Hu-2006}
T. Hu \textit{et al.}, Biophys. J. \textbf{90}, 2731 (2006).

\bibitem{Calladine}
C. R. Calladine \textit{et al.}, {\it Understanding DNA, the
molecular \& how it works} (Elsevier Academic Press, 2004).

\bibitem{Shore}
D. Shore \textit{et al.}, Proc. Natl. Acad. Sci. USA \textbf{78},
4833 (1981).

\bibitem{Cloutier04}
T. E. Cloutier and J. Widom, Molecular Cell {\bf 14}, 355 (2004);
Proc. Natl. Acad. Sci. USA {\bf 102}, 3645 (2005).

\bibitem{Vologodskii05}
Q. Du \textit{et al.}, Proc. Natl. Acad. Sci. USA {\bf 102}, 5397
(2005).

\bibitem{Yuan08}
C. Yuan \textit{et al.}, Phys. Rev. Lett.
{\bf 100}, 018102 (2008).

\bibitem{Shimada}
J. Shimada, H. Yamakawa, Macromolecules, \textbf{17}, 689 (1984).

\bibitem{ZC-j}
Y. Zhang and D. M. Crothers, Biophys. J. {\bf 84} 136 (2003).

\bibitem{Marko04}
J. Yan and J. F. Marko, Phys. Rev. Lett. {\bf 93}, 108108 (2004).

\bibitem{Ranjith05}
P. Ranjith \textit{et al.}, Phys. Rev. Lett. {\bf 94}, 138102
(2005).

\bibitem{Nelson05}
P. A. Wiggins \textit{et al.}, Phys. Rev. E {\bf 71}, 021909 (2005).

\bibitem{Cocco05}
N. Douarche and S. Cocco, Phys. Rev. E {\bf 72}, 061902 (2005).

\bibitem{Wiggins06}
P. A. Wiggins and P. C. Nelson, Phys. Rev. E {\bf 73}, 031906 (2006).

\bibitem{Spakowitz06}
A. J. Spakowitz, Europhys. Lett. {\bf 73}, 684 (2006).

\bibitem{Popov07}
Y. O. Popov and A. V. Tkachenko, Phys. Rev. E {\bf 76}, 021901 (2007).

\bibitem{Frey2007}
K. Alim and E. Frey, Phys. Rev. Lett. {\bf 99}, 198102 (2007); Eur.
Phys. J. E {\bf 24}, 185 (2007).

\bibitem{lankas06}
F. Lankas \textit{et al.}, Structure {\bf 14}, 1527 (2006).

\bibitem{sarah}
S. A. Harris \textit{et al.}, Nucl. Acids Res. {\bf 36}, 21 (2008).

\bibitem{sarah2}
T. B. Liverpool \textit{et al.}, Phys. Rev. Lett. {\bf 100}, 238103
(2008).

\bibitem{maha}
A. Balaeff \textit{et al.}, Phys. Rev. Lett. {\bf 83}, 4900 (1999).

\bibitem{Farshid-03-05}
F. Mohammad-Rafiee and R. Golestanian, Eur. Phys. J. E {\bf 12}, 599 (2003);
J. Phys. Condens. Matter {\bf 17}, S1165 (2005).

\bibitem{Farshid-PRL05}
F. Mohammad-Rafiee and R. Golestanian, Phys. Rev. Lett. {\bf 94}, 238102 (2005).


\bibitem{Marko94}
J. F. Marko and E. D. Siggia, Macromolecules {\bf 27}, 981 (1994);
{\bf 29}, 4820(E) (1996).


\bibitem{Rudnick99}
B. Fain, J. Rudnick, Phys. Rev. E {\bf 60}, 7239 (1999);
R. D. Kamien, Rev. Mod. Phys. {\bf 74}, 953 (2002).

\bibitem{lankas}
F. Lankas \textit{et al.}, Biophys. J. {\bf 85}, 2872 (2003).

\bibitem{LanLif}
L. D. Landau and E. M. Lifshitz, {\em Statistical Physics}, Part 1
(Butterworth-Heinemann, Oxford, 1980).

\bibitem{Lankas2000}
F. Lankas \textit{et al.}, J. Mol. Biol. {\bf 299}, 695 (2000);
S. Neukirch, Phys. Rev. Lett. {\bf 93}, 198107 (2004);
V. Rossetto, Europhys. Lett. {\bf 69}, 142 (2005).

\bibitem{White69}
J. H. White, Am. J. Math. {\bf 91}, 693 (1969).

\end{thebibliography}
\end{document}